# The Atacama Large Millimeter/submillimeter Array: Overview & status

Anthony J. Beasley, Richard Murowinski, Massimo Tarenghi,
Atacama Large Millimeter/submillimeter Array,
Joint ALMA Office,
El Golf 40, Piso 18, Las Condes, Santiago, Chile.

## ABSTRACT

The Atacama Large Millimeter/submillimeter Array (ALMA) is an international radio telescope under construction in the Atacama Desert of northern Chile. ALMA will be situated on a high-altitude site at 5000 m elevation which provides excellent atmospheric transmission over the instrument wavelength range of 0.3 to 3 mm. ALMA will be comprised of two key observing components – an array of up to sixty-four 12-m diameter antennas arranged in a multiple configurations ranging in size from 0.15 to ~14 km, and a set of four 12-m and twelve 7-m antennas operating in closely-packed configurations ~50m in diameter (known as the Atacama Compact Array, or ACA), providing both interferometric and total-power astronomical information. High-sensitivity dual-polarization 8 GHz-bandwidth spectral-line and continuum measurements between all antennas will be available from two flexible digital correlators.

At the shortest planned wavelength and largest configuration, the angular resolution of ALMA will be 0.005". The instrument will use superconducting (SIS) mixers to provide the lowest possible receiver noise contribution, and special-purpose water vapor radiometers to assist in calibration of atmospheric phase distortions. A complex optical fiber network will transmit the digitized astronomical signals from the antennas to the correlators in the Array Operations Site Technical Building, and post-correlation to the lower-altitude Operations Support Facility (OSF) data archive. Array control, and initial construction and maintenance of the instrument, will also take place at the OSF. ALMA Regional Centers in the US, Europe and Japan will provide the scientific portals for the use of ALMA; a call for early science observations is expected in 2009. In this paper, we present the status of the ALMA project as of mid 2006.

## 1. INTRODUCTION

ALMA has been designed to provide sensitive spectra and images in the wavelength range from 0.3 to 3 mm of atomic & molecular gas, thermal & non-thermal electrons and thermal dust in our Solar System, the Galaxy, nearby galaxies and high-redshift universe. These data will provide new and unique insights into the formation of galaxies, stars, planets and the chemical precursors necessary for life itself. ALMA will complement 8-10 meter optical/near-IR telescopes such as the Very Large Telescope, Gemini, Subaru and also the Hubble Space Telescope and its successor, the James Webb Space Telescope, with its ability to image dust enshrouded objects or cold molecular material.

Three exciting new observing challenges have been used to define the primary technical specifications of ALMA: (1) the ability to detect spectral line emission from rotational spectral lines of the carbon monoxide molecule, atomic and ionized carbon in a galaxy with the properties of the Milky Way at a redshift of $z=3$ in less than 24 hours of measurement; (2) to image the kinematics of gas in protostars and protoplanetary disks around young solar type stars out to a distance of 500 light years (this represents the distance to the nearby well-known clouds in Ophiuchus, Taurus or Corona Australis); and (3) to provide high-fidelity precise images at an angular resolution better than 0.1".

ALMA's flexible design will support:

- Imaging the broadband emission from dust in evolving galaxies at epochs of formation as early as $z=10$;

- Tracing the chemical composition of star-forming gas in galaxies throughout the history of the universe through measurements of molecular and atomic spectral lines;

- Measuring the motions of obscured galactic nuclei and Quasi-Stellar Objects on spatial scales finer than 300 light years;
- Imaging and spectroscopy of gas-rich heavily obscured regions that are collapsing to form protostars, protoplanets and pre-planetary disks;
- Measuring the crucial isotopic and chemical gradients within circumstellar shells that reflect the chronology of stellar nuclear processing ;
- Producing sub-arcsecond images of cometary nuclei, hundreds of asteroids, Centaur and Kuiper belt objects together with images of planets and their moons;
- Observations of active solar regions to investigate particle acceleration on the Sun's surface.

For more complete descriptions of ALMA science see Wilson (2005) or Wootten (2001), or visit www.alma.info.

## 2. OVERVIEW

ALMA is a partnership between Europe, Japan and North America in cooperation with the Republic of Chile. In Europe it is funded by the European Southern Observatory (ESO), in Japan by the National Institutes of Natural Sciences (NINS) in cooperation with the Academia Sinica in Taiwan, and in North America by the US National Science Foundation (NSF), in cooperation with the National Research Council of Canada (NRC). ALMA construction and operations are carried out on behalf of Europe by ESO, on behalf of Japan by the National Astronomical Observatory of Japan (NAOJ) and on behalf of North America by the National Radio Astronomy Observatory (NRAO), which is managed by Associated Universities, Inc. (AUI). ALMA project development is coordinated by the Joint ALMA Office (JAO), based in Santiago, Chile.

The high-level technical specifications for the ALMA are shown in Table 1. ALMA will be built on the Chajnantor altiplano in the Atacama Desert of northern Chile at an elevation of slightly over 5000 m. The site is administered by the Chilean Ministry of National Assets and set aside by Presidential decree as a protected region for science. Measurements made since 1995 of the atmospheric transparency and stability confirm that the site has superior conditions for millimeter and submillimeter-wavelength astronomy.

The ALMA antennas each have a primary reflecting surface 12 meters in diameter with a parabolic cross-section. The materials used in their construction have been selected to allow the antennas to maintain their performance when fully exposed to the thermal variations and wind gusts imposed by the site environment. Each antenna is fully steerable, and more than 85 percent of the celestial sphere is above the horizon at the Chajnantor site. The antennas can be moved (reconfigured) among 186 prepared antenna locations to provide a range of spatial resolutions in the final astronomical images. Each station has a concrete foundation to support the antenna and provision for electrical power and fiber-optic based data communications. The antennas are moved by a pair of specially-designed rubber-tired antenna transporters currently under construction. ALMA will utilize a range of antenna configurations, forming arrays as small as 150 meters in diameter (for the study of large or low surface brightness objects) and as large as 14.5 km in diameter (for the study of small, high surface brightness objects). The ACA's four 12 m and twelve 7 m antennas will be located in a more compact configuration ~50-m in diameter to allow sensitive wide-field imaging and total power measurements. Three different manufacturers are involved in producing antennas for ALMA: in Europe, a consortium of Alcatel Alenia Space, EIE & MT Aerospace; in the US, Vertex RSI; and the Mitsubishi Electric Company in Japan. During 2002-2005 these companies built prototype antennas at the NRAO Very Large Array site in New Mexico to demonstrate their technical ability to meet the demanding ALMA antenna technical specifications.

Each antenna will be equipped initially with a receiving system (Front End) capable of detecting astronomical signals in seven wavelength bands. The design and infrastructure of ALMA will allow the installation of up to ten receiver bands, eventually covering all the millimeter/submillimeter atmospheric transmission windows from 9 mm to 0.3 mm. The ALMA receivers are coherent detectors, meaning they strictly preserve phase across the elements of the array. To achieve this, a common local oscillator signal is distributed to all antennas to convert the received astronomical signal to a much lower intermediate frequency that is transmitted to a central high-site technical building, where it is correlated with the signals from all other antennas. Each receiver cartridge includes two receivers which operate in orthogonal linear polarizations to allow the complete polarization state of received signals to be measured, and have been implemented using all-electronic receiver tuning. All receivers utilize superconducting mixers that operate at temperatures at ~4 K. The receivers for each antenna are housed in a common cryogenic dewar located at the Cassegrain focus of each antenna.

## Table 1: ALMA Technical Summary

### Array

| | |
|---|---|
| **Number of Antennas (N)** | Up to 64 + 16 *(ACA)* |
| **Total Collecting Area ($\pi/4\ ND^2$)** | 7238 m$^2$ + 913 m$^2$ *(ACA)* |
| **Angular Resolution** | 0.2" $\lambda$ (mm)/baseline (km) |

### Array Configurations

| | |
|---|---|
| **Compact: Filled area diameter** | 150 m + ~50m *(ACA)* |
| **Maximum Baseline (weighted)** | 14.5 km |
| **Total Number of Antenna Stations** | 186 + 22 *(ACA)* |

### Antennas

| | |
|---|---|
| **Diameter (D)** | 12 m, 12 & 7 m *(ACA)* |
| **Surface Accuracy** | 25 microns RMS |
| **Pointing** | 0.6" RMS in 9 m/s wind |
| **Path Length Error** | < 15 microns during sidereal track |
| **Fast Switch** | 1.5 degrees in 1.5 secs, 1.8 secs *(ACA)* |
| **Total Power** | **Instrumented and gain stabilized** |
| **Transportable** | **Special-purpose vehicle on rubber tires** |

### Front Ends

| | |
|---|---|
| 84 -116 GHz | |
| 125 – 163 GHz | *All receivers:* |
| 211 - 275 GHz | *- Dual polarization* |
| 275 - 370 GHz | *- Noise performance limited* |
| 385 – 500 GHz | *- SIS* |
| 602 - 720 GHz | |
| *787 – 950 GHz* | *(planned: NAOJ)* |
| | |
| **Water Vapor Radiometer** | 183 GHz |

| Signal Transmission | |
|---|---|
| Bandwidth | 8 GHz, each polarization |
| IF Transmission | Digital (digitized at antennas) |
| Local Oscillator | Photonic |

| Correlator | |
|---|---|
| Correlated Baselines | 2016 + 120 *(ACA)* |
| Bandwidth | 16 GHz per antenna |
| Spectral Channels | 4096 per IF |

| Data Rate | |
|---|---|
| Data Transmission from Antennas | 120 Gb/s per antenna, continuous |
| Signal Processing at the Correlator | $1.6 \times 10^{16}$ multiply/add per second |

Also mounted at the Cassegrain focus is a water vapor radiometer tuned to the 183 GHz line of terrestrial water emission. These devices will be used to correct for the atmospheric phase distortions caused by fluctuations in the amount of water vapor over the site, which would otherwise seriously limit the performance of the array over long baselines and/or shorter wavelengths. Amplitude calibration of ALMA will use a multiple-temperature load system and standard astronomical flux-density benchmarks.

The received signals are amplified, digitized at the antenna, and returned to the control building via fiber optics connections. In order to process the 16 GHz bandwidth IF, digital electronics subdivides that signal into eight 2 GHz sub-bands for transmission to the correlator. Timing signals and reference oscillators synchronize the operation of the antennas and the data collection. Buried power and fiber optic connections link each station to the Technical Building at the Array Operations Site (AOS-TB), which houses the array correlator and support electronics (including the LO system, fiber patch panel and computers) and contains an interim control room for array operations and hardware testing.

The astronomical signals are processed in the AOS-TB by a correlator: a special-purpose digital signal processor. It combines the digitized IF signals from all the antennas pair-wise and produces a set of complex correlation coefficients (fringe amplitude and phase) as a function of baseline and frequency. Images of the angular distribution of the radio emission from the astronomical source on the sky are created by Fourier inversion of these complex (phase and amplitude) data. For these recycling correlators the product of total bandwidth with number of channels is a constant. For a 2 GHz bandwidth, two polarizations, the correlator provides 128 spectral channels for each of the 2016 + 120 baselines correlated. The finest frequency channel will be 6 kHz, or 0.02 km s$^{-1}$ at 100 GHz.

To support the construction, maintenance and operation of ALMA, an Operations Support Facility (OSF) is under construction 24 km from the AOS-TB at a 2900m elevation site. The OSF provides a pleasant working environment for staff involved in a broad range of activities. Scientific operation of the array will be from a control room at the OSF via a high speed digital link to the AOS-TB. Infrastructure at the OSF will consist of the antenna service building, array control building, electronic laboratories, and office, administrative and residential facilities. The OSF is connected to the AOS by a road constructed to transport the antennas and the operations/maintenance staff.

The ALMA computing system has the task of scheduling observations on the array, controlling all the array instruments, including pointing the antennas, monitoring instrument performance, monitoring environmental parameters, managing the data flow through the electronics, and presentation of these data to the correlator. The correlator output must be processed through an image pipeline, where it is calibrated and first-look images produced. Finally, the science data and all associated calibration data, monitor data, and derived data products are archived and made available for network transfer. In full operation, the standard output from ALMA will be calibrated images and spectra that have been processed in a standard set of reduction programs linked in a pipeline. The user will receive these images, together with the correlated data (*uv*-data files), calibration files, and monitor information files. The average data rate is expected to be ~6.4 MB/s, with a peak projected rate an order of magnitude higher. These results will also be stored in a data archive and delivered to the astronomers in a timely manner. An office in Santiago will house ALMA administrative and local scientific staff and support the Chilean astronomy community's use of ALMA. Additional support facilities in North America, Europe & Japan (the ALMA Regional Centers, or ARCs) will provide interfaces and user support between the instrument and their regional astronomical communities. Further calibration, image processing and analysis of the astronomical data will be carried out at the ARCs. A diverse community will use and benefit from ALMA's powerful scientific capabilities, and producing an easy-to-use system for both novices and experts is a key system design goal.

## 3. STATUS

All major elements of ALMA development are progressing smoothly, including:

- Site: construction of the AOS-TB and OSF facilities is underway, with completion expected late 2007. The transporter-capable road from the OSF to the AOS is complete, and worker and contractor accommodation camps at the OSF are in use. Construction of the AOS antenna stations will begin late 2006.

- Antennas: production antenna contracts for ALMA were signed during the past year with US, European & Japanese firms; the first production antenna is expected Q1 2007, the final deliveries in late 2011. The first transporter is expected mid 2007.

- Receivers & electronics: the challenging task of building the electronics systems is proceeding well, with debugging of the digital transmission systems underway and testing of first production receiver anticipated early 2007 at the ALMA Test Facility site in New Mexico.

- Correlators: These are under construction; the first quarter of the 64-antenna correlator is complete, and the second underway; the ACA correlator is under construction in Japan.

- Computing: the distributed computing team is producing software for all aspects of array testing, commissioning, operations and offline data reduction.

Major upcoming milestones for the project include:

- First production antenna accepted for start of assembly, integration & verification tasks: Quarter 1 2007
- Completion of the AOS-TB and OSF buildings: Q4 2007
- Two-antenna interferometry at the OSF: Q2 2008; three-antenna interferometry at the AOS: Q3 2008
- Call for Early Science proposals from the community: Q1 2009
- Delivery of main-array antenna #32: Q2 2011; #50 in late 2011.
- Start of full operations: Q3 2012.

*This paper is presented by the JAO on behalf of the scientists, engineers, technical and administrative staff of the bilateral ALMA Project and our colleagues from ALMA-J (NAOJ).*

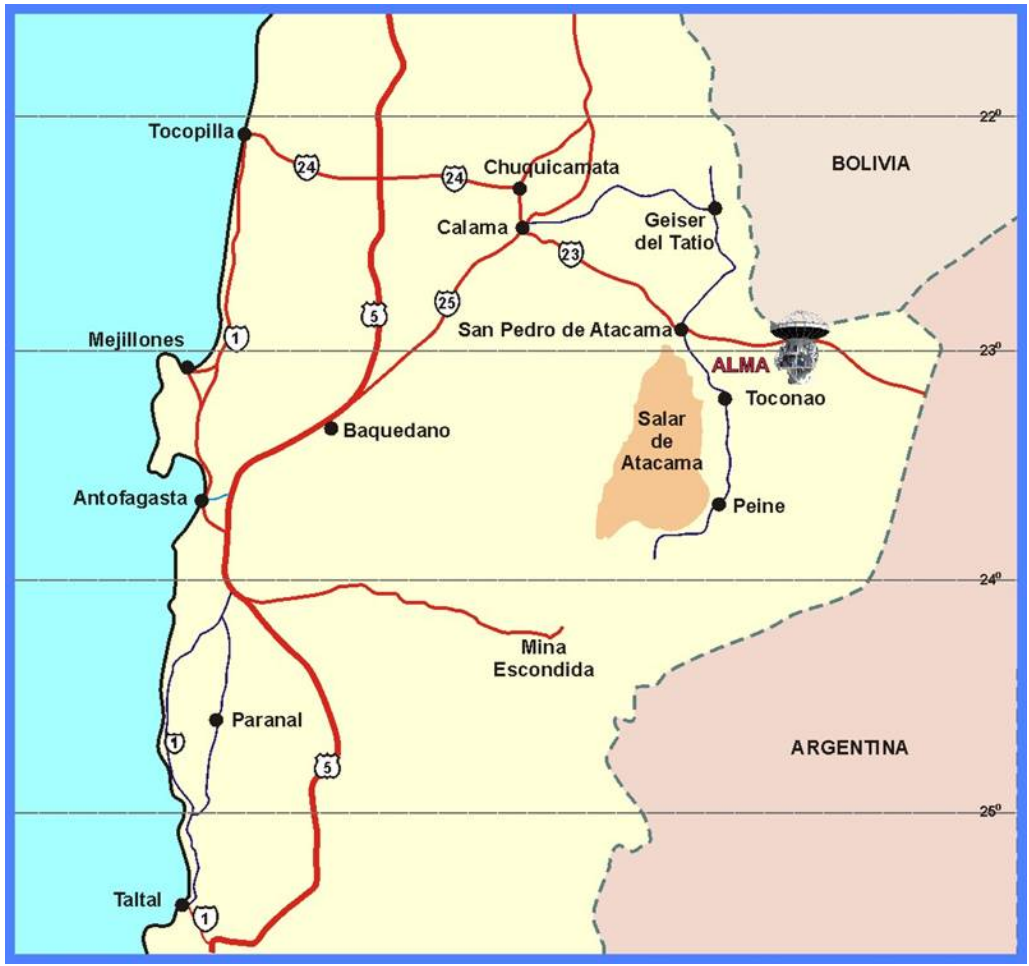

Fig. 1. Location of the ALMA array site near San Pedro de Atacama, northern Chile.

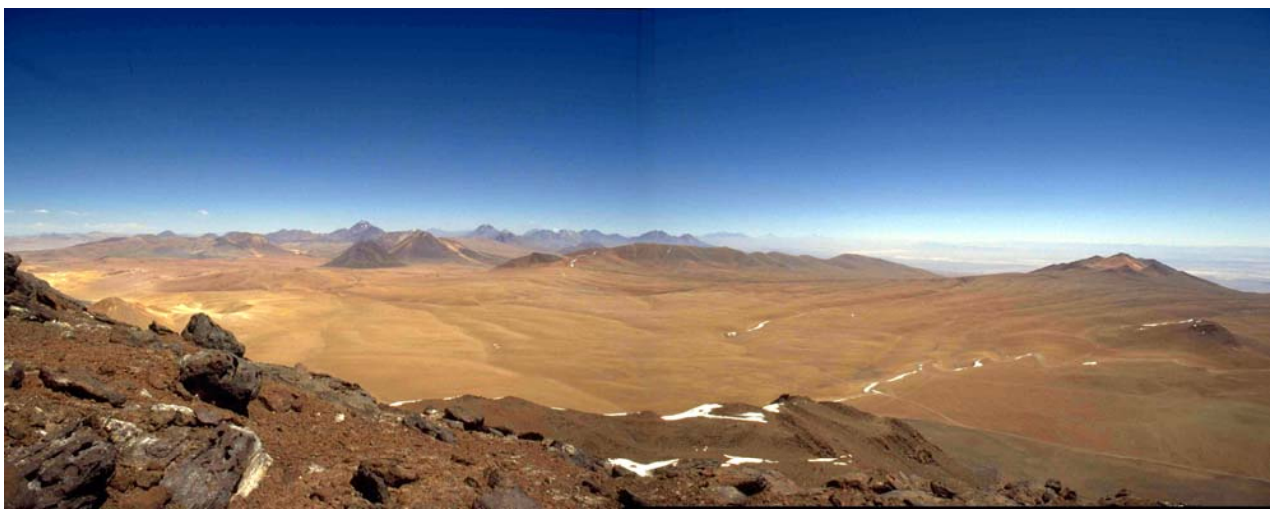

Fig.2. The ALMA site at 5000 m *(photo: S. Radford)*.

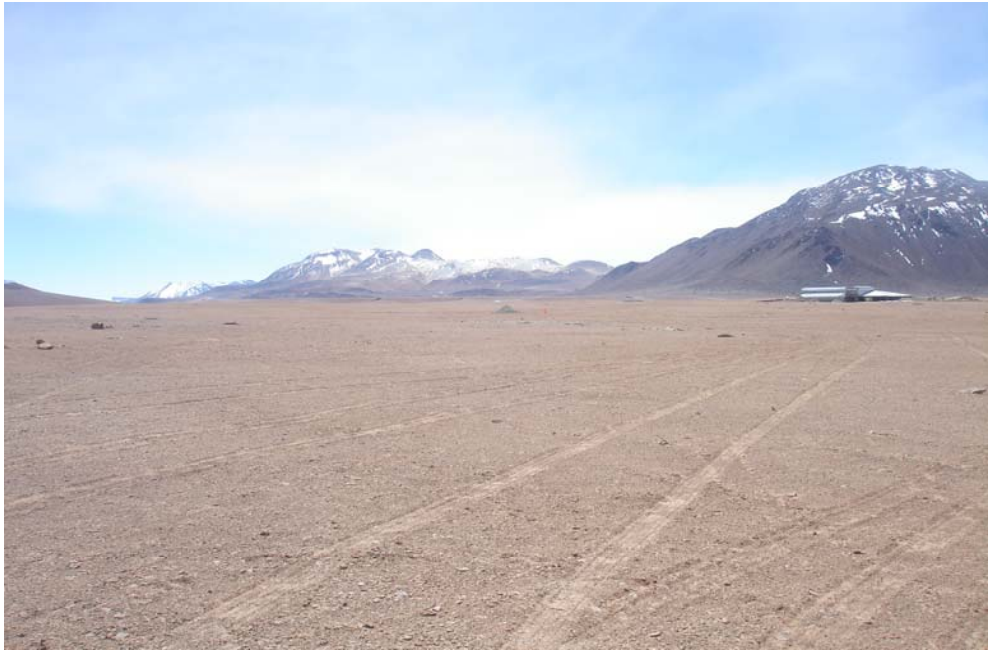

Fig.3. The center of the area planned for antenna stations; the AOS-TB is visible to the right.

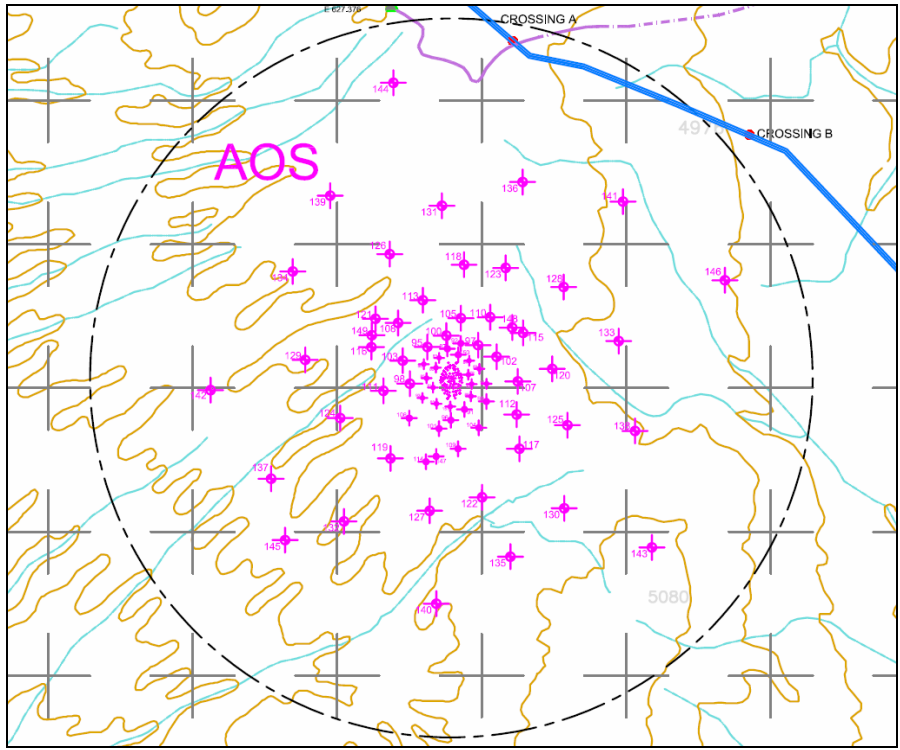

Fig.2. The antenna locations at the AOS. The circle shown is 5 km in diameter. A network of roads will continuously move antennas throughout the configuration of antenna stations. A larger "Y+" configuration spreading the antennas out over a ~14-18 km area has also been designed.

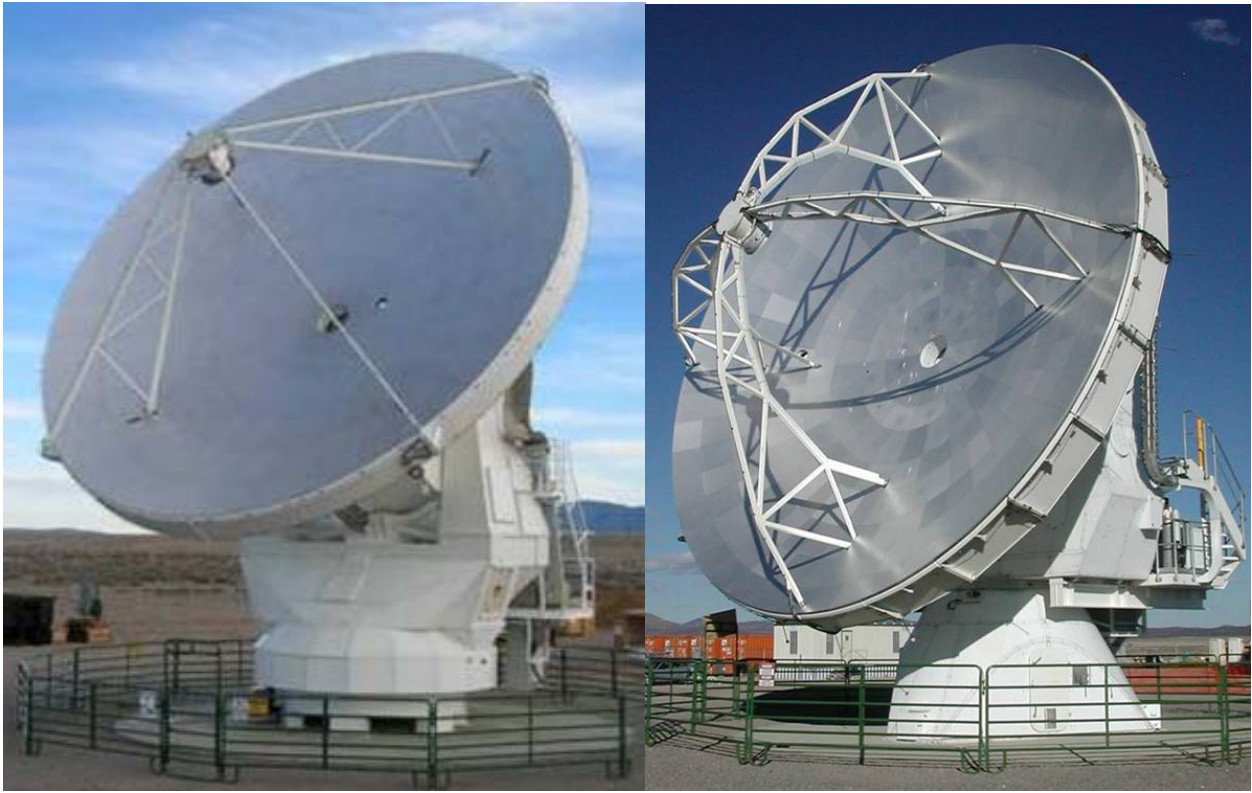

Fig.6. The Alcatel Consortium (left) and Vertex (right) antenna prototypes, built at the VLA site in New Mexico.

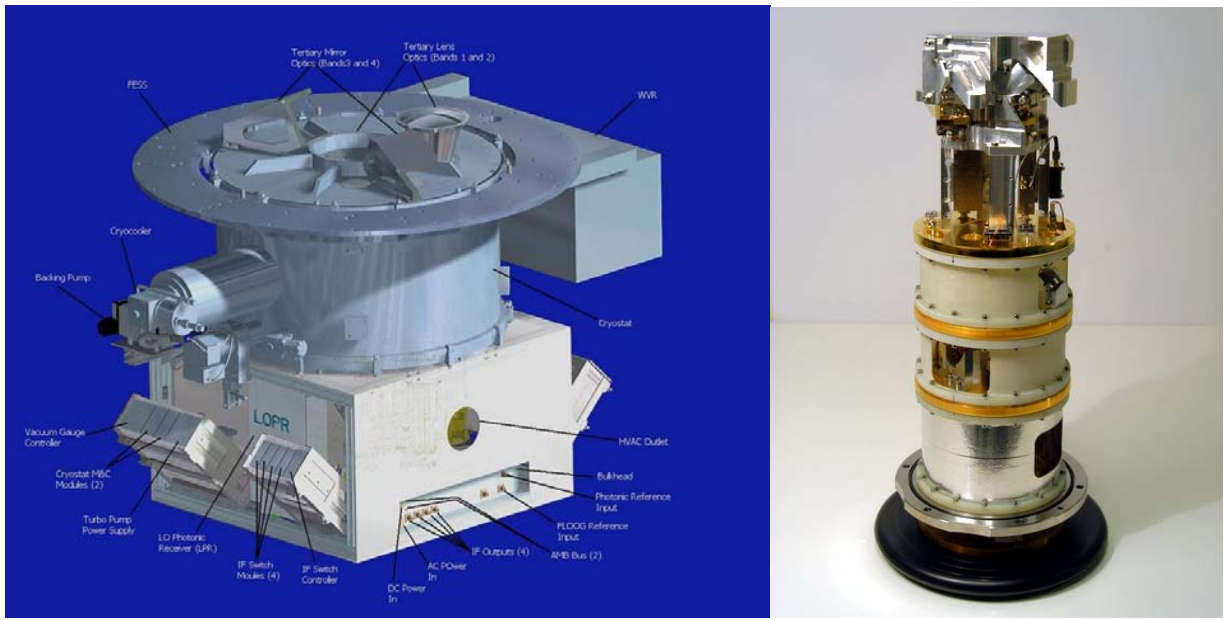

Fig.7. The Front End cryostat holding the antenna receivers (left), and a 600 GHz receiver cartridge (right).

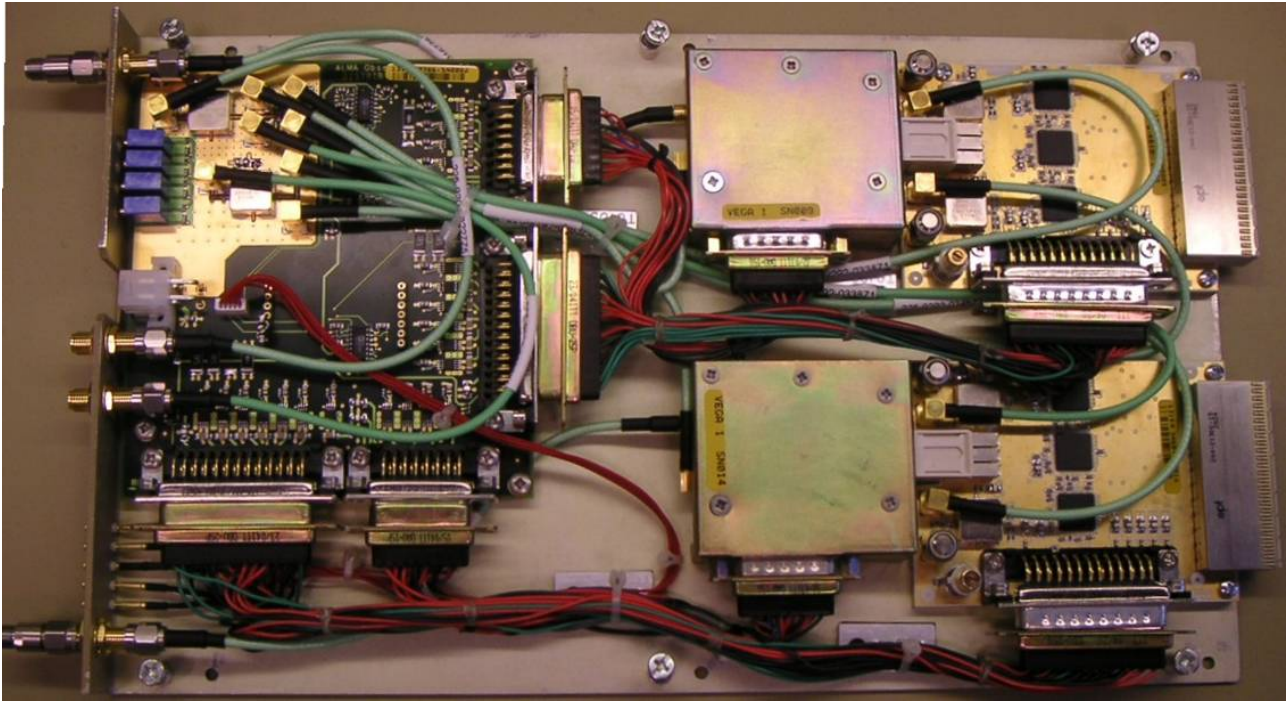
Fig.8. A University of Bordeaux high-speed digitizer developed for ALMA.

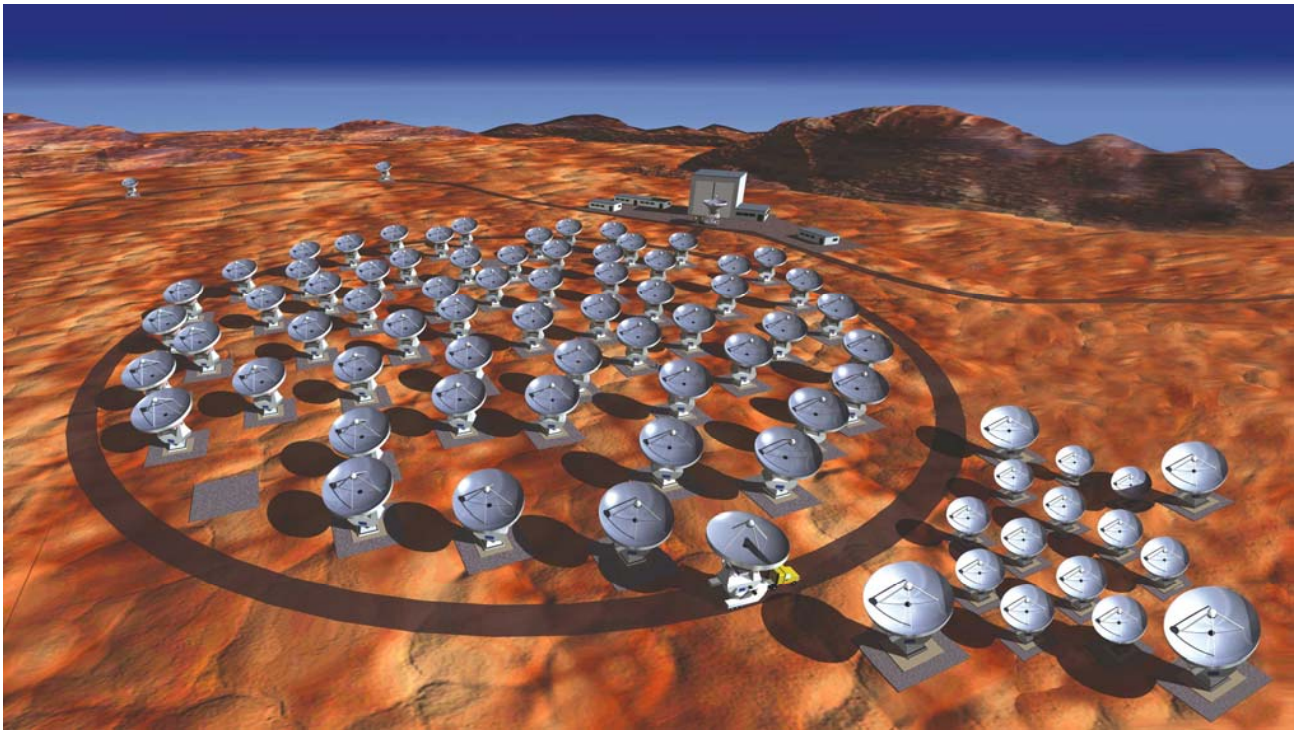
Fig.9. Simulation of ALMA 64-antenna array (left) plus ACA (bottom right) on the Chajnantor site.